# Large-scale nonlinear optical computing with incoherent light via linear diffractive systems

Alexander Chen[1,†], Yuntian Wang[1,2,3,†], Md Sadman Sakib Rahman[1,2,3], Yuhang Li[1,2,3], Aydogan Ozcan[1,2,3,*]

**Affiliations**

[1]Electrical and Computer Engineering Department, University of California, Los Angeles, CA, 90095, USA

[2]Bioengineering Department, University of California, Los Angeles, CA, 90095, USA

[3]California NanoSystems Institute (CNSI), University of California, Los Angeles, CA, 90095, USA

*ozcan@ucla.edu    [†] Equal contribution

**Abstract**

Nonlinear computation is essential for various information-processing tasks. Optical implementations are attractive because passive light propagation can manipulate high-dimensional signals with extreme throughput and parallelism; yet realizing nonlinear mappings in optical hardware remains challenging due to the weak nonlinearity of optical materials and the large intensities required to induce nonlinear interactions. This challenge is further amplified in many systems that operate with incoherent illumination, motivating a coherence-aware framework for scalable optical nonlinear processing. Here, we show that linear optical systems—in particular, optimized diffractive processors comprising passive surfaces—can perform large-scale nonlinear function approximation under spatially incoherent or partially coherent illumination, when preceded by intensity-only input encoding. We quantify how the accuracy of the nonlinear function approximation varies with the degree of parallelism, the number of diffractive layers, and the number of trainable diffractive features. Numerical results demonstrate snapshot computation of up to one million distinct nonlinear functions in a single forward pass through a diffractive processor, with the function outputs spatially multiplexed and read out using densely packed detectors at the output. We further provide a proof-of-concept experimental demonstration under incoherent illumination from a liquid crystal display (LCD), enabled by a model-free *in situ* learning strategy that jointly optimizes the diffractive profile and detector readout geometry in the presence of hardware imperfections and misalignments. Our findings establish diffractive processors as a massively parallel universal function approximator for both spatially incoherent and partially coherent illumination.

# Introduction

Computing has been pivotal in scientific discovery and technological progress, with digital electronics prevailing due to their scalability and precision. However, the slowdown of conventional device scaling and the rapid growth of machine-learning workloads have intensified interest in alternative computing architectures that can deliver higher throughput per unit energy(*1–3*). In particular, the rapid expansion of modern large-scale artificial intelligence (AI) models has heightened concerns about the energy footprint of computation and the supporting infrastructure, motivating sustained efforts toward efficient, domain-specific hardware accelerators(*4, 5*). Optical implementations are attractive in this context because passive wave propagation can manipulate high-dimensional signals with extreme parallelism and low latency(*6–9*). This advantage is especially relevant when the information of interest is already carried by light, for example, in phase imaging(*10*), microscopy(*11*), sensing(*12*), and machine vision(*13*)—where optical frontends can provide high-throughput analog preprocessing of information carried by light prior to electronic readout and digital post-processing.

These motivations have driven increasing interest in photonic and neuromorphic optical computing platforms across both free-space and integrated photonics-based implementations(*14–26*). For example, diffractive optical processors have emerged as a versatile paradigm for task-specific analog computing(*7, 27–35*). Architecturally, these systems comprise cascaded, spatially structured passive diffractive surfaces interspersed with free-space propagation. Data-driven, task-specific optimization enables these diffractive processors to manipulate optical fields with high fidelity, successfully approximating various linear(*28, 36–38*) and nonlinear(*39–46*) functions using structured, passive linear optical materials. Alternative free-space optical computing strategies that leverage nonlinear materials generally require optical intensities that are prohibitive for practical applications(*47–49*); optical field-enhancement structures can be used to partially mitigate this limitation at the cost of introducing additional complexity(*50*), optical losses(*51*), or bandwidth constraints(*52*). Most of these free-space diffractive designs are based on coherent illumination sources; however, various applications demand incoherent illumination and a passive diffractive processor for universal nonlinear computation with incoherent light has not yet been explored. Consequently, developing a *coherence-aware* strategy that realizes high-throughput nonlinear processing under incoherent or partially coherent illumination represents an important next frontier.

Here, we introduce a diffractive computing framework for massively parallel nonlinear function approximation that operates under spatially incoherent illumination and extends naturally to partially coherent illumination. Built upon an intensity-only input encoding strategy, our approach circumvents the strict coherence requirements of conventional designs by synthesizing a large set of desired nonlinear functions through the interplay of an intensity-based input encoding, spatially varying intensity point spread functions and square-law detection at the output plane. We report a statistical optical forward model that explicitly accounts for partial coherence of illumination via ensemble averaging over independent field realizations, enabling the optimization of passive phase-only processors that maintain function approximation accuracy under phase fluctuations. Leveraging these optimized designs, we numerically demonstrate the one-shot parallel approximation of up to one million distinct nonlinear functions in a single forward pass through a passive optical processor, with all the spatially multiplexed nonlinear function outputs read out in parallel using a detector array at the output plane. We also present a proof-of-concept experimental validation under incoherent LCD illumination using a model-free *in situ* learning strategy that jointly optimizes the diffractive phase profile and detector readout geometry directly on hardware, i.e., without the knowledge of the optical forward model.

Validated through both numerical and experimental analyses, this work positions passive diffractive processors composed of linear materials as a scalable, highly parallel nonlinear function approximator for optical analog computing under varying degrees of spatial coherence.

**Results**

First, we demonstrate massively parallel optical approximation of a set of $N_f$ distinct nonlinear functions by synergizing a predetermined input encoding strategy with an optimizable linear optical processor under spatially incoherent illumination. Figure 1 schematically illustrates the proposed framework for massively parallel nonlinear function approximation. As shown in Fig. 1A, the input function argument $a$ is first encoded into an intensity-only pattern; the intensity values within this pattern correspond to the cosine and sine bases in the Fourier representation of the target nonlinear functions. The spatially incoherent illumination, upon modulation by this encoding pattern (representing the input function argument $a$), is processed by an optimized $K$-layer diffractive processor that is fixed; this diffractive processor produces an output intensity pattern that corresponds to the target nonlinear functions at different output detectors, each of which is assigned to a distinct nonlinear function. All the diffractive layers, following their in-silico optimization for a given set of $N_f$ nonlinear functions, remain fixed during inference at any value of the input function argument $a$. At the output plane, a tile of $2 \times 2$ photo-detectors is designated for each target function, $f_i$. A differential combination of the four measured intensities within each tile yields the corresponding function output, as shown in Fig. 1B; integrating $N_f$ such tiles at the output plane enables the simultaneous readout of $N_f$ target functions – executed all in parallel through a static/fixed diffractive optical processor. Therefore, for each input argument $a$, this diffractive framework produces $N_f$ (e.g., $10^6$) nonlinear function outputs in one shot, simultaneously implementing $N_f$ nonlinear mappings of $a$ with low latency, through a thin diffractive optical volume that axially spans $< 5\sqrt{N_f}\lambda$, where $\lambda$ is the illumination wavelength (see Methods for details).

The input encoding strategy is further detailed in Fig. 1C, where the intensity distribution serves as the controllable quantity at the input plane, representing, on demand, the varying values of the input function argument $a$. Specifically, the input argument $a$ is mapped to $2N_p$ intensity values at the input plane, i.e., $I_{\cos}(p; a) = (1 + \cos(2\pi p a))/2$ and $I_{\sin}(p; a) = (1 + \sin(2\pi p a))/2$, for $p = 1, \cdots, N_p$. In other words, $N_p$ denotes the number of harmonics in the Fourier representations of the target nonlinear functions of interest.

It is important to note that these Fourier frequencies should *not* be confused with the spatial frequencies of the input light carried by propagating waves. Unlike free-space or lens-based

Fourier processors that harness spatial frequencies carried by light—all of which represent spatially *invariant* convolutional systems—the optimized diffractive processor in our case has spatially varying incoherent point spread functions, enabling it to perform, all in parallel, multiple distinct nonlinear input–output mappings that are spatially multiplexed under spatially incoherent or partially coherent light. The harmonics that are represented at the input encoding plane of a diffractive processor are not related to the spatial frequencies at the input aperture, and they correspond to the harmonics that synthesize the nonlinear output functions through an optimized set of spatially varying intensity point spread functions.

According to our architecture shown in Fig. 1, the encoded intensity-only input pattern comprises $2N_p$ pixels, $N_p$ for the cosine basis components and $N_p$ for the sine basis components of the desired nonlinear functions to be optically implemented. Without loss of generality, we assume the scalar input $a \in [-0.5, 0.5]$, since any finite input range can be affinely rescaled to this interval. An arbitrary nonlinear target function $f_i(a)$ can be approximated by a truncated Fourier expansion,

$$f_i(a) \approx C_i + \sum_{p=1}^{N_p} \left(A_{i,p} \cos(2\pi p a) + B_{i,p} \sin(2\pi p a)\right), i = 1,2,\ldots,N_f. \quad (1)$$

where $C_i$ represents the constant component of each nonlinear function of interest that is independent of the input argument, $a$. Without loss of generality, our analysis focuses on normalized 1D nonlinear functions that satisfy an output range of $[0, 1]$. Therefore, during the training and testing of a diffractive processor for nonlinear function approximation, we normalize each output function without altering its shape (see Methods for details).

Under spatially incoherent illumination, the diffractive optical processor implements a linear transformation of the input intensity, corresponding to a set of spatially varying intensity point spread functions(*53, 54*). Therefore, if we collect the $2N_p$ intensity values in the encoded intensity pattern into a vector $\mathbf{i}(a) \in \mathbb{R}^{2N_p}$, the all-optical approximation output $\hat{f}_i(a)$ can then be written as,

$$\hat{f}_i(a) = \mathbf{w}_i^T \mathbf{i}(a), i = 1,2,\ldots,N_f, \quad (2)$$

where $\mathbf{w}_i \in \mathbb{R}^{2N_p}$ is determined by the trained diffractive layers. Given that the elements of $\mathbf{i}(a)$ are nonnegative sinusoidal functions, i.e., $(1 + \cos(2\pi p a))/2$ and $(1 + \sin(2\pi p a))/2$, we get,

$$\hat{f}_i(a) = \hat{C}_i + \sum_{p=1}^{N_p} \left(\hat{A}_{i,p} \cos(2\pi p a) + \hat{B}_{i,p} \sin(2\pi p a)\right), i = 1,2,\ldots,N_f, \quad (3)$$

where $\hat{A}_{i,p}$ and $\hat{B}_{i,p}$ are specifically determined by $\mathbf{w}_i$. Consequently, the diffractive processor performs nonlinear function approximation by learning the effective Fourier coefficients $\{\hat{A}_{i,p}, \hat{B}_{i,p}\}$ which are mapped into spatially varying intensity point spread functions channeling the harmonic terms at the input encoding pixels into different output detectors, each assigned to different nonlinear functions; as a result of this multiplexed mapping, the diffractive optical processor can execute, all in parallel, all the members of $\hat{f}_i(a)$ for $i = 1, 2, \ldots, N_f$. To enable this, the diffractive layers are trained/optimized such that, for each function $f_i$ and harmonic index $p$, the optically realized coefficients satisfy $\hat{A}_{i,p} \approx A_{i,p}$ and $\hat{B}_{i,p} \approx B_{i,p}$. In this optimization process, the constant offset $\hat{C}_i$ does not need to be explicitly enforced since both the target function and the synthesized optical readout are normalized with an output range of $[0, 1]$. Consequently, the corresponding DC component automatically satisfies $\hat{C}_i \approx C_i$, leading to $\hat{f}_i(a) \approx f_i(a)$ for an optimized diffractive processor. In our implementation, $\hat{f}_i(a)$ is decoded from its designated $2 \times 2$ detector tile via a differential combination of the four measured intensities (as shown in Fig. 1B) with weights $\begin{bmatrix} +1 & +1 \\ -1 & -1 \end{bmatrix}$. This differential readout strategy(55) enables the approximation of Fourier series terms with negative weights, while the row-wise summation improves robustness by averaging pixel-level measurements; this also means only $2N_f$ detectors are needed at the output plane. Figure 1D illustrates the successful approximation of various target nonlinear functions, demonstrating a strong correspondence between the diffractive processor's outputs and the theoretical ground truth across selected values of the scalar input argument $a$.

Next, we report the performance of three different diffractive optical processors, independently optimized to perform, in a snapshot, $N_f = 10{,}000$, $N_f \approx 100{,}000$ and $N_f = 1{,}000{,}000$ nonlinear function approximations, respectively; see Fig. 2. In this analysis, we set the total number ($N$) of trainable diffractive features to $N \approx 2N_iN_o$, distributed evenly across $K = 4$ diffractive surfaces. Here, $N_i = 2N_p$ refers to the number of pixels at the input aperture used for encoding, with $N_p = 9$ for all the settings, unless otherwise stated. At the output plane, each distinct function is assigned a $2 \times 2$ detector tile; after row-wise summation (top-row sum and bottom-row sum), each tile provides 2 independent measurements, the difference of which corresponds to the function output, as detailed in the earlier paragraph. Therefore, we have the effective output dimension as $N_o = 2N_f$. Substituting $N_i = 2N_p$ and $N_o = 2N_f$ yields $N \approx 8N_pN_f$. Consequently, the number of trainable diffractive features for each optical processor design scales linearly with $N_f$ for a fixed encoding bandwidth, $N_p$. During the inference of nonlinear function outputs, spatial incoherence is emulated by Monte Carlo averaging over random illumination phase patterns: for each input intensity, we performed coherent propagation over $N_{\phi,\text{test}} = 20{,}000$ independent random phase realizations at the diffraction limit of light, and averaged the resulting output intensities (see Methods for details).

The per-function mean squared error (MSE) values between the target functions ($f_i(a)$) and the optical diffractive output approximations ($\hat{f}_i(a)$) across the entire set of functions ($i = 1, 2, \ldots, N_f$) are calculated and reported in Fig. 2A; the results reveal that the MSE distributions for $N_f = 10{,}000$ and $N_f \approx 100{,}000$ are tightly concentrated in the order of $10^{-4}$, while scaling to $N_f = 1{,}000{,}000$ increases the error as expected. Nevertheless, the overall error level remains remarkably low (in the order of $10^{-3}$). Representative approximations in Fig. 2B further support this trend: for $N_f = 10{,}000$ and $N_f \approx 100{,}000$, the median-case examples show no visually discernible mismatch between the target functions and the diffractive output approximation, whereas at $N_f = 1{,}000{,}000$ the discrepancies become more noticeable.

Having validated the framework's efficacy in accommodating massive functional parallelism, the investigation next proceeds to the architectural configuration underpinning these transformations. First, the effect of the number of diffractive layers is analyzed under a fixed number of trainable features, $N$. Using the same configuration as in Fig. 2, the total number of trainable diffractive features remains constant at $N \approx 8N_p N_f$ and the number of diffractive layers varies ($K = 1, 2$, and $4$). The resulting optimized phase profiles of the diffractive surfaces are shown in Fig. 3A. Compared with the single-layer design ($K = 1$) that has all the $N$ diffractive features on a single diffractive layer, the multi-layer optical processors ($K = 2$ and $4$) exhibit phase modulation over a noticeably larger effective active area (visible as the expanded central regions), whereas the peripheral area in the $K = 1$ case lacks structure and contributes little to the diffractive output. This trend suggests that an increased number of diffractive layers enables more efficient use of the available degrees of freedom (DoF) for the desired multi-function nonlinear approximation. The corresponding spatial map of function approximation errors at the output plane, as shown in Fig. 3B, further highlights the benefit of an increased number of diffractive surfaces: increasing from $K = 1$ to multi-layer designs reduces the overall error scale by nearly two orders of magnitude, shifting from $\sim 10^{-1}$ for $K = 1$ to $\sim 10^{-3}$ for $K = 2$ and $4$. Collectively, these findings suggest that distributing a constant budget of trainable diffractive features ($N$) across multiple layers significantly boosts the representational capacity for massively parallel nonlinear function approximation, highlighting an architectural depth advantage.

Building on this observation of depth-enhanced feature utilization, we next analyze the intrinsic trade-off between the trainable feature budget ($N$) and the resulting approximation accuracy. Figure 4 further characterizes the relationship between $N$ and the function approximation accuracy, and quantifies the benefit of increasing the number of diffractive layers with the same DoF. For this analysis, we defined a feature sufficiency ratio $r = N/(8N_p N_f)$ and reported the average per-function MSE across all target functions together with the worst-case performances (i.e., the MSEs of the worst 1% and the worst 5% of the functions in terms of approximation errors). For the multi-layer designs ($K = 2$ and $4$), increasing $r$ consistently lowers both the average and worst-case errors, with the most pronounced gains observed at small $r$ and a clear tendency to saturate as $r$ approaches 1. In contrast, the single-layer architecture ($K = 1$) exhibits substantially larger errors across all metrics and a much higher error floor, indicating that a shallow diffractive architecture severely limits nonlinear function approximation even when additional trainable diffractive features are provided. Notably, while the approximation performance degrades for $r < 1$, the marginal gains from $r = 1.0$ to $1.3$ are minimal. It thereby confirms that $N \sim 2N_i N_o = 8N_p N_f$ (corresponding to $r \approx 1$) provides the requisite degrees of freedom to realize nonlinear transformations with negligible error.

We next examine the spectral limits of nonlinear function approximation under spatially incoherent illumination. The complexity of nonlinear functions supported by the proposed diffractive framework is governed by the encoding bandwidth, $N_p$, i.e., the number of cosine/sine bases used to encode the input. Larger $N_p$ provides a Fourier representation with more harmonics, thereby enabling the approximation of higher-bandwidth functions with more rapid variations. Although the proposed architecture is not specifically designed as an optical analogue of neural network activation functions, such nonlinear functions defined on a finite input interval can still be emulated within the same diffractive framework. Figure 5 demonstrates an accurate approximation of common activation functions used in deep learning and highlights the role of the input encoding dimension, $N_p$. We encoded the scalar input argument $a$ using cosine and sine intensity patterns with $N_p$ harmonics (Fig. 5A) and allocated separate output regions to compute four target activation functions all in parallel—ReLU, Sigmoid, Tanh, and Softplus (Fig. 5B). As shown in Fig. 5C, the diffractive optical outputs closely track the ground truth functions for all four nonlinear activation functions, and the MSE decreases systematically as $N_p$ increases from 9 to 49 and 100. Because these nonlinear activation functions each have different output ranges, we affinely normalized each target function to $[0, 1]$ during the training and testing, and computed the MSE values in this normalized domain to ensure a consistent error scale across different nonlinear functions; the plotted curves in Fig. 5 are then mapped back to their original output ranges for visualization. The accuracy improvement with larger $N_p$ is most pronounced for functions with sharp transitions (e.g., ReLU) and steep slopes (e.g., scaled Sigmoid/Tanh/Softplus), indicating that finer input encoding provides the high-frequency basis required for high-fidelity nonlinear function approximation. In these cases, the MSE values are computed over $a \in [-0.5 + \epsilon, 0.5 - \epsilon]$ with $\epsilon = 0.025$ to mitigate Gibbs phenomenon-induced boundary artifacts in the error estimate; the relationship between the approximation accuracy and $\epsilon$ is reported in fig. S1.

Most practical illumination sources are partially coherent, and the effective coherence diameter of an illumination beam increases linearly with the axial distance between the emitter and the object aperture of interest, motivating a generalization of our framework under partially coherent illumination (see Fig. 6). To design partially coherent diffractive nonlinear function approximators, the detected output intensity is computed by averaging over an ensemble of independent field realizations characterized by spatial phase patterns with a certain correlation length that statistically determines the spatial coherence diameter of the illumination. As shown in Fig. 6A, the scalar function input $a$ is encoded into an intensity-only pattern and propagated through the optimized $K$-layer diffractive processor; in this design, we used $K = 4$, $N_p = 9$ and $N = 4 \times 200 \times 200$. Each one of the $N_f = 100$ target nonlinear functions is decoded from a dedicated detector tile using the same differential combination of the measured intensities, same as in the spatially incoherent case. In this proof-of-concept analysis, we set the phase correlation length to $C_\phi \approx 2.1\lambda$; see the Methods section. Under this partially coherent illumination, the optical diffractive output remains in close agreement with the target nonlinear functions. Representative best-, median-, and worst-case examples in Fig. 6B show accurate tracking of the target nonlinear functions across the input interval. The function approximation MSE distribution in Fig. 6C also indicates that the target functions are reconstructed with consistently low errors, and the spatial error map in Fig. 6D shows no pronounced localization of inaccuracies in specific regions of the output field of view, as desired. Together, these results demonstrate that the proposed diffractive design retains its massively parallel nonlinear function approximation capability even when the illumination is partially coherent.

Having established numerically that the proposed framework retains robust nonlinear-function approximation capability under incoherent and partially coherent illumination, we next turn to a proof-of-concept experimental validation in the visible spectrum. As shown in Fig. 7A-7B, the experimental setup comprises an LCD, relay optics, a phase light modulator (PLM), and a complementary metal-oxide-semiconductor (CMOS) camera. The LCD was used for intensity-only input encoding under incoherent red illumination (~0.62 μm), the PLM functioned as the trainable diffractive processor to modulate the optical field, and the CMOS image sensor captured the output intensity patterns. Because of the limited optical power available from the LCD, two relay lenses were used to improve optical throughput: the first lens established the LCD and PLM as conjugate planes, and the second lens guided the reflected PLM field onto the CMOS sensor plane. Since we used a model-free *in situ* learning strategy (detailed in the Methods section), the knowledge of the parameters of this experimental system, including the axial distances and the illumination wavelength, was not used.

Using this hardware platform, we experimentally validated nonlinear function approximation under incoherent LCD illumination for $N_f = 4$. The in situ-learned optimized phase profile is shown in Fig. 7C after deployment onto the PLM. Figure 7D presents representative intensity distributions measured at the CMOS plane, overlaid with the final selected positive and negative detector regions that define the differential readout for each target nonlinear function (see the Methods section). As reported in Fig. 7E, the experimental optical outputs remain in good agreement with the ground-truth target functions across four distinct random nonlinear mappings, yielding an average MSE of $1.06 \times 10^{-2}$, successfully capturing the overall functional trends. To obtain these results in the presence of misalignments, imperfections, and measurement noise, we developed an adaptive detector-pruned *in situ* learning strategy, illustrated in Fig. 8, that jointly optimizes the diffractive phase profile and the detector readout geometry directly on the hardware platform without knowledge of the optical forward model. The procedure begins with 200 randomly initialized differential detector pairs for each target nonlinear function and progressively removes persistently underperforming detector pairs, thereby suppressing readout regions affected by, e.g., optical dead zones and/or hardware-induced artifacts; see the Methods section for details. After this pruning stabilizes, the surviving top-$k$ detector pairs for each nonlinear function are subjected to additional *in situ* fine-tuning, where $k$ denotes the number of highest-performing retained detector pairs passed to this refinement stage. In the final stage, each retained differential pair is decoupled into its positive and negative detector regions, and a combinatorial selection step determines the most robust readout configuration. This multi-stage optimization procedure yields the sparse detector layout shown in Fig. 7D and enables reliable nonlinear-function approximation under incoherent illumination despite the non-idealities of the physical system.

**Discussion**

We introduced a diffractive optical computing framework for massively parallel nonlinear function approximation under spatially incoherent and partially coherent illumination. By combining an intensity-only encoding of the input argument $a$ with an optimized multi-layer diffractive processor, this architecture simultaneously evaluates up to $N_f = 10^6$ distinct nonlinear functions in a single optical forward pass using incoherent or partially coherent light. Importantly, the time-of-flight through this thin diffractive volume (axially spanning $< 5\sqrt{N_f}\lambda$) enables low latency for snapshot massively parallel nonlinear approximations with minimal electronic post-processing and no optical material nonlinearity. These results establish a scalable route toward large-scale incoherent and partially coherent optical computing, where the computational throughput is fundamentally tied to the space–bandwidth products of the input encoding plane ($2N_p$) and the output readout plane ($2N_f$). In addition, we provided a proof-of-concept experimental validation under incoherent LCD illumination using an adaptive detector-pruned *in situ* learning strategy that is forward model-free, demonstrating that the framework can be translated to physical hardware despite misalignments, imperfections and measurement noise.

At the core of our approach is the observation that, under spatially incoherent illumination, a diffractive processor with sufficiently large $N$ and architectural depth can be optimized to implement any arbitrary set of spatially varying intensity point spread functions between an input aperture and an output aperture(*38, 53*). We exploit this property by encoding the input argument $a$ into nonnegative cosine and sine intensity channels that span the space of Fourier-representable functions, subsequently learning an optical transformation that simultaneously synthesizes a large set of desired nonlinear functions at the output plane using these learned spatially varying intensity point spread functions. While this strategy conceptually resembles feature-space methods in machine learning—using passive optics for physical feature mapping—it faces a fundamental physical constraint: optical intensity is real and non-negative. Because the Fourier series expansion of an arbitrary target function $f(a)$ generally relies on a superposition of weighted terms where the weights can be negative, the incoherent diffractive processor requires a mechanism to represent negative values. To practically realize these signed Fourier coefficients using strictly non-negative intensity measurements, we implemented a differential readout scheme at the output plane. In this configuration, introduced in the Results section, each target function is assigned to a dedicated detector tile, and the scalar output is decoded via a differential combination of the measured intensities. This approach not only provides the necessary degrees of freedom to approximate the target function over the normalized output domain but also improves robustness through pixel averaging (see Fig. 1B). Furthermore, this multiplexed readout concept can be extended beyond real-valued scalar functions; for example, additional detectors can be allocated to separately encode the real and imaginary parts when complex-valued nonlinear function approximation under spatially incoherent or partially coherent illumination is desired.

In terms of the training strategy, we compared two training objectives for optimizing the same diffractive architecture and output readout (see fig. S2). Under the $\mathcal{L}_1$ objective (see the Methods for details), the diffractive model is supervised on the final scalar output obtained from the differential combination of the $2 \times 2$ detector tile per nonlinear function (Fig. 1B), providing an *end-to-end learning* that directly matches each target function. In fig. S2, we considered an alternative training objective, $\mathcal{L}_2$, in which supervised learning is applied to the even and odd readouts separately prior to their combination. These two formulations are mathematically equivalent at the level of the final readout since any function can be uniquely decomposed into an even and an odd function; however, the component-wise objective implicit in $\mathcal{L}_2$ splits the gradients into two coupled subtasks that must be satisfied simultaneously by the same set of diffractive features. In practice, gradients associated with the even and odd functions can compete for limited representational capacity and partially cancel each other through shared optical pathways, slowing function-specific optimization and increasing sensitivity to readout imbalance and tile-level crosstalk. Consistent with this intuition, our simulations reported in Supplementary Fig. 2 show that direct end-to-end supervision with the $\mathcal{L}_1$ objective yields a statistically significant improvement over $\mathcal{L}_2$, along with more accurate function approximations and lower spatially distributed errors across the output field of view, validating the effectiveness of the end-to-end training objective set by $\mathcal{L}_1$ (see the Methods).

Our results and analyses also shed light on how the accuracy of nonlinear function approximation scales with the trainable feature budget. We showed that increasing the number of trainable diffractive features improves the mean per-function MSE, while an insufficient number of features leads to a rapid degradation in performance. Moreover, increasing the number of diffractive layers provides a strong and consistent benefit: for a fixed $N$, multi-layer designs achieve substantially lower approximation errors than a single-layer diffractive processor with the same $N$, indicating that the multi-layer, deeper architecture of a diffractive processor enhances representational efficiency for simultaneously implementing the desired set of nonlinear functions at scale. From a degree-of-freedom perspective, our analysis also suggests that the complexity of the approximation problem scales with the product of the input encoding ($N_p$) and the number of nonlinear functions ($N_f$), effectively exhibiting a first-order dependence, $O(N_p N_f)$. If one could independently control both the amplitude and phase of each diffractive feature (rather than the phase-only modulation considered here), the required number of optimized features could be reduced at the cost of output photon efficiency, suggesting potential benefits of more expressive optical modulation primitives.

While our demonstrations focus on universal approximation of randomly generated nonlinear functions, the framework naturally supports structured function families. As shown with some of the commonly used nonlinear activation functions reported in Fig. 5, increasing the number of harmonics $N_p$ at the input plane systematically improves the approximation fidelity for functions with sharper transitions or higher effective bandwidth. Fundamentally, the framework functions as a universal approximator under spatially incoherent and partially coherent light; provided an adequate harmonic basis and trainable diffractive parameters, continuous nonlinear functions over a bounded domain can be reconstructed to a desired accuracy using linear optical propagation coupled with intensity detection at the output plane.

We also considered a practically important regime where illumination is neither fully coherent nor fully incoherent. By modeling partial coherence through averaging over a finite number of independent field realizations and controlling the phase correlation length, we observed that an optimized diffractive design retains its nonlinear function approximation capability across the output field of view.

Beyond the numerical robustness observed under partial coherence, the proof-of-concept experiments further showed that hardware deployment can benefit from jointly optimizing the diffractive phase profile and detector readout geometry through adaptive *in situ* learning, which helps suppress localized readout failures caused by experimental imperfections, misalignments and noise - all in a model-free manner.

Looking ahead, scaling the parallelism toward far larger $N_f$ (e.g., tens of millions of nonlinear functions) will benefit from advances in large-area, high-resolution diffractive layer fabrication and 3D alignment, as well as high-pixel-count image sensors and compact optical packaging. Additional future directions include extending the functional encoding to multi-dimensional inputs $\vec{a} = (a_1, a_2, ...)$, exploring alternative feature bases beyond Fourier harmonics for improved efficiency on specific target function classes, and co-designing diffractive processors with task-level objectives to enable end-to-end optical preprocessing and massively parallel analog computation under spatially incoherent or partially coherent illumination.

**Methods**

**Forward model of spatially incoherent and partially coherent diffractive processors**

To model diffractive processors under spatially incoherent/partially coherent illumination, we used coherent-field propagation as a building block. The optical output of incoherent/partially coherent illumination is then obtained via time-averaging of coherently propagated outputs over randomly fluctuating input phases. The diffractive processor is modeled as a cascade of $K$ phase-only diffractive surfaces separated by free-space propagation distances $\{\Delta z_\ell\}_{\ell=0}^{K}$, where $\ell = 0$ denotes propagation from the input/encoding plane to the first diffractive layer, and $\ell = K$ denotes propagation from the last diffractive layer to the output/detector plane. The complex optical field immediately after the $\ell$-th diffractive layer is:

$$U_\ell^+(x,y) = U_\ell^-(x,y)\, T_\ell(x,y), \quad \ell = 1,2,\ldots,K, \tag{4}$$

where $U_\ell^-(x,y)$ and $U_\ell^+(x,y)$ denote the incident and exiting fields at the $\ell$-th layer, respectively, and $T_\ell(x,y) = \exp(j\phi_\ell(x,y))$ is the learnable phase-only transmittance with $\phi_\ell(x,y) \in [0, 2\pi)$.

Free-space propagation between successive diffractive planes is modeled using the angular spectrum (AS) method(56). Let $\mathcal{F}\{\cdot\}$ and $\mathcal{F}^{-1}\{\cdot\}$ denote the 2D Fourier transform pair with spatial-frequency coordinates $(f_x, f_y)$. Then, propagation over a distance of $\Delta z$ is given by:

$$U(x,y; z+\Delta z) = \mathcal{F}^{-1}\{\mathcal{F}\{U(x,y;z)\} \cdot H_{AS}(f_x, f_y; \Delta z)\}, \tag{5}$$

where $H_{AS}(f_x, f_y; \Delta z)$ is the free-space transfer function for the propagation distance of $\Delta z$, given by

$$H_{AS}(f_x, f_y; \Delta z) = \begin{cases} \exp\left\{j\dfrac{2\pi}{\lambda}\Delta z \sqrt{1 - (\lambda f_x)^2 - (\lambda f_y)^2}\right\}, & f_x^2 + f_y^2 < \dfrac{1}{\lambda^2} \\ 0, & f_x^2 + f_y^2 \geq \dfrac{1}{\lambda^2} \end{cases}, \tag{6}$$

where $\lambda$ is the illumination wavelength.

The axial spacing between adjacent diffractive planes is set to $\Delta z_\ell = \Delta z = W\sqrt{(2\delta/\lambda)^2 - 1}$, where $W = \sqrt{N/K}\,\delta$ denotes the lateral width of the diffractive surface, $N$ is the total number of trainable diffractive features, $K$ is the number of diffractive surfaces and $\delta$ is the lateral feature size at each diffractive layer. The illumination wavelength and feature size are set to $\lambda = 550$ nm and $\delta = 300$ nm, respectively; these choices determine the effective numerical aperture (NA) between the diffractive layers to be ~0.75.

With spatially incoherent or partially coherent illumination, the average output optical intensity $O(x,y)$ of a diffractive processor for a given input intensity $I(x,y)$ is computed by averaging the coherent output intensities over random input phase realizations:

$$O(x,y) = \langle |\mathcal{D}\{\sqrt{I(x,y)}\exp(j\phi^{(r)}(x,y))\}|^2 \rangle_r = \lim_{N_\phi \to \infty} \frac{1}{N_\phi} \sum_{r=1}^{N_\phi} |\mathcal{D}\{\sqrt{I(x,y)}\exp(j\phi^{(r)}(x,y))\}|^2, \quad (7)$$

where $\mathcal{D}\{\cdot\}$ denotes coherent propagation of the optical field through the diffractive processor, and $\langle \cdot \rangle$ denotes the statistical average over independent phase realizations $\phi^{(r)}(x,y)$, which approximates the measured intensity for a sufficiently long exposure time. Unless otherwise specified, the number of random phase realizations used during each inference is set to $N_{\phi,test} = 2 \times 10^4$. For the approximation of the nonlinear activation functions shown in Fig. 5, we used $N_{\phi,test} = 10^6$ to accurately quantify the impact of increased $N_p$ on the function approximation performance. For spatially incoherent illumination, the input phase is modeled as i.i.d. (independent and identically distributed) random variables:

$$\phi_{\text{incoh}}(x,y) \sim \text{Uniform}[0, 2\pi). \quad (8)$$

For partially coherent illumination, correlated phase masks are generated by first sampling a random field $W(x,y) \sim N(\mu, \sigma_0)$, and then smoothing it with a 2D zero-mean Gaussian kernel $G_\sigma$:

$$\phi_{\text{part}}(x,y) = \text{mod}\left(\frac{2\pi}{\lambda}(W * G_\sigma)(x,y), 2\pi\right), \quad (9)$$

where "$*$" denotes 2D convolution operation, and

$$G_\sigma(x,y) = \exp\left(-\frac{x^2+y^2}{2\sigma^2}\right). \quad (10)$$

The phase correlation length $C_\phi$ for partially coherent illumination is controlled by $\sigma$ and is estimated by fitting to the phase-autocorrelation function $R_\phi(x,y)$ that is defined by(*57*):

$$R_\phi(x,y) = \exp\left(-\frac{\pi(x^2+y^2)}{C_\phi^2}\right). \quad (11)$$

In the simulation of partially coherent illumination, we used $\mu = 25\lambda$, $\sigma_0 = 8\lambda$ and $\sigma = 4\lambda$ for random phase generation, which results in an average correlation length of $C_\phi \approx 2.1\lambda$.

For spatially incoherent propagation, the forward model is accelerated by using an intensity point spread function (PSF)-based approach, in which a linear transformation (matrix) is used to link the vectorized input and output intensity distributions(*38, 53*). The input intensity distribution is modeled as a discrete array of pixel values with a spatial dimension of $\sqrt{N_p} \times 2\sqrt{N_p}$, and is vectorized into $\mathbf{i}(a) \in \mathbb{R}^{2N_p}$ via a lexicographical flattening operation; similarly, the detector intensity is modeled as an array with dimensions $2\sqrt{N_f} \times 2\sqrt{N_f}$, and is vectorized as $\mathbf{o}(a) \in \mathbb{R}^{4N_f}$. For a given diffractive optical processor, the propagation of input intensities under spatially incoherent illumination is given by a linear transformation:

$$\mathbf{o}(a) = \mathbf{H}\mathbf{i}(a), \tag{12}$$

where $\mathbf{H} \in \mathbb{R}^{4N_f \times 2N_p}$ denotes the effective intensity-domain transfer matrix, representing a desired set of spatially varying intensity point spread functions between the input and output apertures. The $k$-th column of $\mathbf{H}$ is determined by computing the vectorized output intensity response resulting from a Kronecker delta function input at the $k$-th pixel coordinate of $I(x, y)$.

**Experimental setup**

The presented nonlinear function approximation framework was experimentally validated under incoherent red illumination from an LCD in the visible spectrum. A liquid crystal display (Elecrow; resolution, $800 \times 480$, pixel pitch, $135\ \mu m$) was used as the illumination source of the system, and only its red illumination channel was utilized. The intensity-encoded pattern on the LCD screen was imaged by a first lens onto a PLM (Texas Instruments, DLP670S; resolution, $2716 \times 1600$, pixel pitch, $5.4\ \mu m$; dual-pixel mode, utilizing $2 \times 2$ neighboring pixels for 4-bit depth phase modulation), which functioned as the trainable diffractive processor. The reflected, modulated optical field from the PLM was subsequently guided by a second lens onto a CMOS camera (Basler GigE acA1920-40gm; resolution, $1920 \times 1200$; pixel pitch, $5.86\ \mu m$). To capture each image, the camera was operated with a fixed exposure time of $250\ ms$. To manage the high dimensionality of the diffractive plane and optimize computational overhead, a pixel binning factor of 2 was applied to the trainable phase parameters before upsampling to the native resolution of the PLM. The resolutions of the LCD screen, the PLM, and the sensor plane were $98 \times 60$, $1600 \times 1600$ and $1920 \times 1200$, respectively.

The target continuous nonlinear mappings ($N_f = 4$) were generated using a random Fourier series expansion defined on a bounded one-dimensional input domain, which was discretized into 16 randomly spaced evaluation points. During the training phase, each scalar input value was dynamically encoded into an intensity-only pattern displayed on the LCD. Following a 0.15-second hardware settling interval, the optical signal propagated to the PLM and then to the camera plane. To realize signed Fourier coefficients using strictly non-negative optical intensity measurements, a spatially multiplexed differential readout scheme was employed(*26, 55*). The system initialized 200 differential detector pairs per target function, randomly positioned across the camera field of view. The measured output for each detector pair was defined as the intensity difference between two co-located integration regions separated vertically by a fixed 10-pixel gap.

Both the target function values and the corresponding optical detector outputs were independently subjected to min–max normalization over the evaluated input interval prior to error computation.

**Training and testing setup**

The diffractive processor models were implemented and trained in PyTorch (v2.9.0) with CUDA 12.8 on a single NVIDIA GeForce RTX 5090 GPU. During training, the model parameters were optimized by backpropagation through a differentiable wave-propagation model using AdamW(*58*), with a learning rate of $10^{-3}$. With this setup, a nonlinear function approximator diffractive model with e.g., $N_f = 10{,}000$, $K = 4$, and $N = 7.2 \times 10^5$ typically converged in ~ 12 hours on an RTX 5090 GPU. The execution time for 100 epochs is ~13 hours on an NVIDIA GTX 1080 Ti GPU, accounting for all system operations such as experimental image capture, data preprocessing, the pruning procedure, and PPO-driven gradient descent.

During each training iteration, we randomly drew a mini-batch of function input arguments $\{a_j\}_{j=1}^{B}$ from the domain $a \in [-0.5,\ 0.5]$, with a batch size of $B = 32$. These arguments were encoded into input intensity patterns and propagated through the diffractive optical processor using the intensity-domain transfer matrix for the spatially incoherent model. For the partially coherent case, we estimated the optical response during training by averaging over $N_{\phi,\text{train}} = 100$ random phase realizations. During the training process, function normalization, $\mathcal{N}\{\cdot\}$, was applied to both the optical outputs and the target nonlinear function within each mini-batch prior to loss computation. We used 1000 mini-batches per epoch, and 750 epochs during each training. For blind testing, we used the Monte Carlo method with $N_{\phi,\text{test}} = 2 \times 10^4$, except for the analysis reported in Fig. 5, which used $N_{\phi,\text{test}} = 10^6$ to better highlight the impact of increased $N_p$ on the function approximation accuracy.

**Supplementary Materials file includes**:

- Figs. S1 to S2.

- Supplementary Methods for:

    o Encoding of the function argument

    o *In situ* learning strategy

    o Training and Testing of Spatially Incoherent and Partially Coherent Diffractive Processors

**Figures**

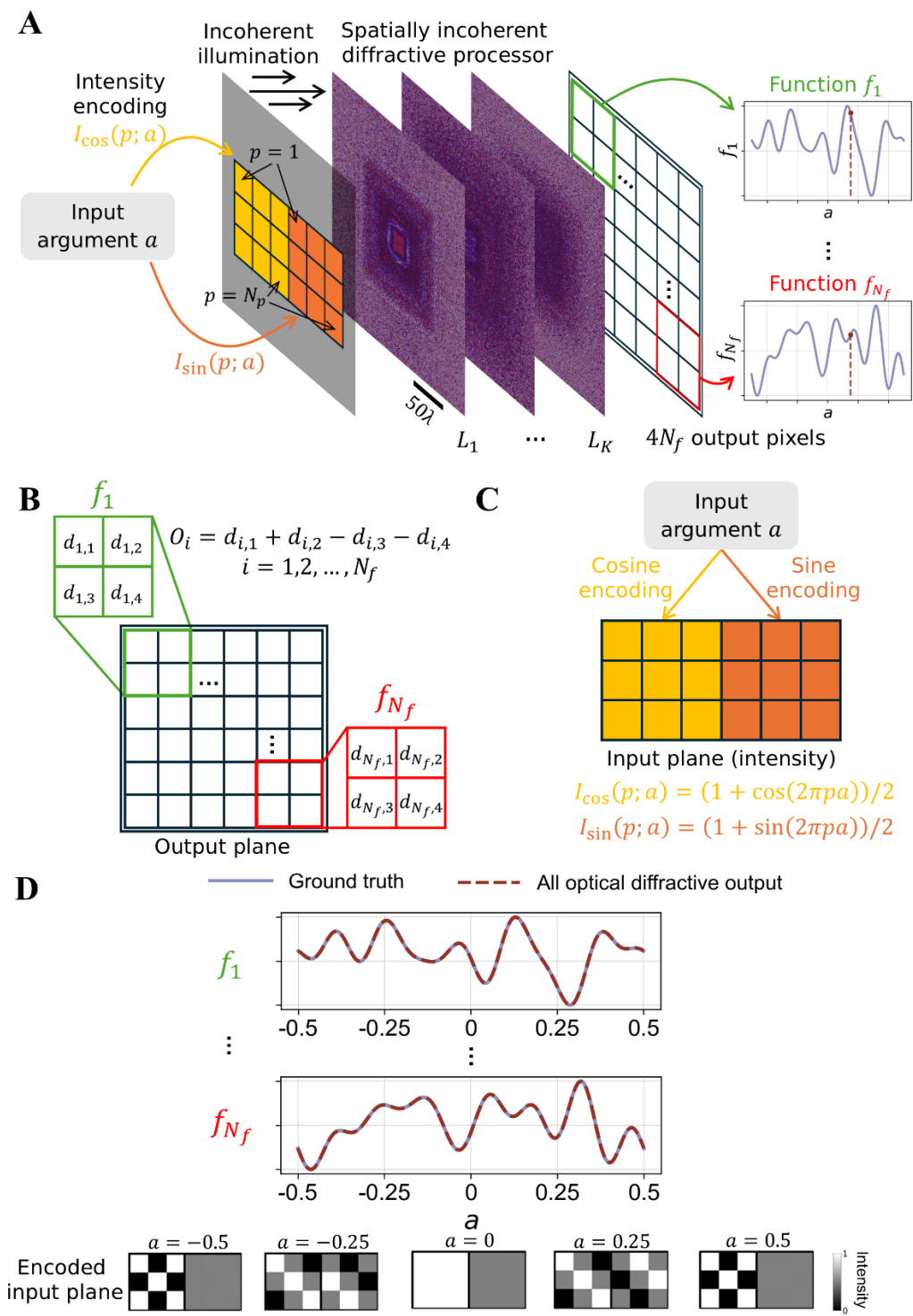

**Fig. 1 Massively parallel nonlinear function approximation with a diffractive processor under spatially incoherent illumination. (A)** Conceptual overview of the framework. The input

argument $a$ is encoded into an intensity-only pattern and propagates through a $K$-layer diffractive processor under spatially incoherent illumination. Intensities measured at the output plane are decoded to yield $N_f$ function values, executed all in parallel. **(B)** Readout at the output plane. Each function $f_i$ is assigned a $2 \times 2$ detector tile, and a differential combination of the measured intensities yields the corresponding scalar output. **(C)** Intensity encoding at the input plane. Cosine and sine intensity blocks form the basis for encoding the scalar input $a$ and approximating an arbitrary set of target nonlinear functions. **(D)** Representative results comparing ground truth functions (solid) with the diffractive outputs (dashed); the bottom row shows encoded input patterns for selected values of $a$.

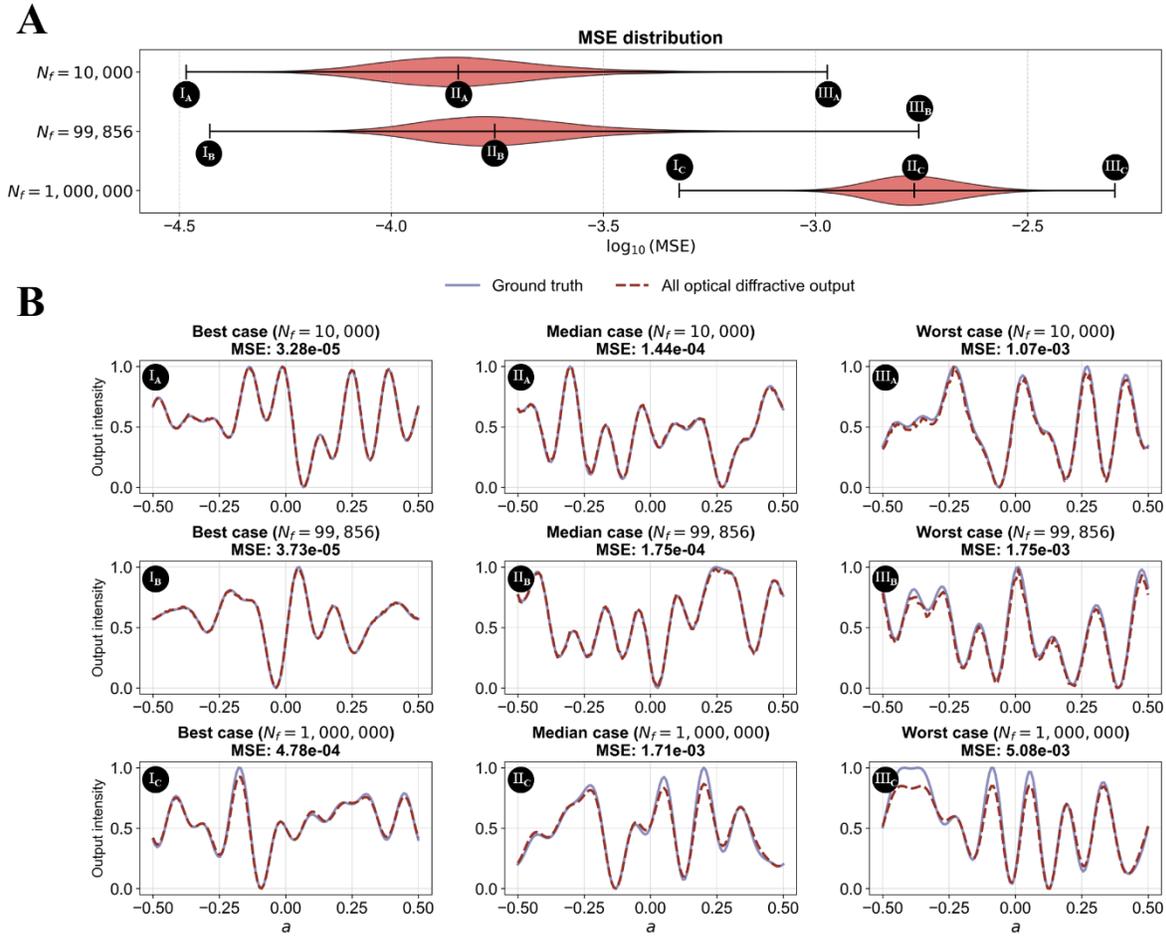

**Fig. 2 Approximation error scaling with the number ($N_f$) of target nonlinear functions. (A)** Distributions of per-function MSE values between the ground truth nonlinear functions and the diffractive outputs for $N_f = 10,000$, $N_f \approx 100,000$, and $N_f = 1,000,000$. Markers indicate the best-, median-, and worst-case functions selected for visualization. **(B)** Representative approximations for the best, median, and worst functions at each $N_f$. Solid curves denote ground truth, and dashed curves denote the diffractive outputs.

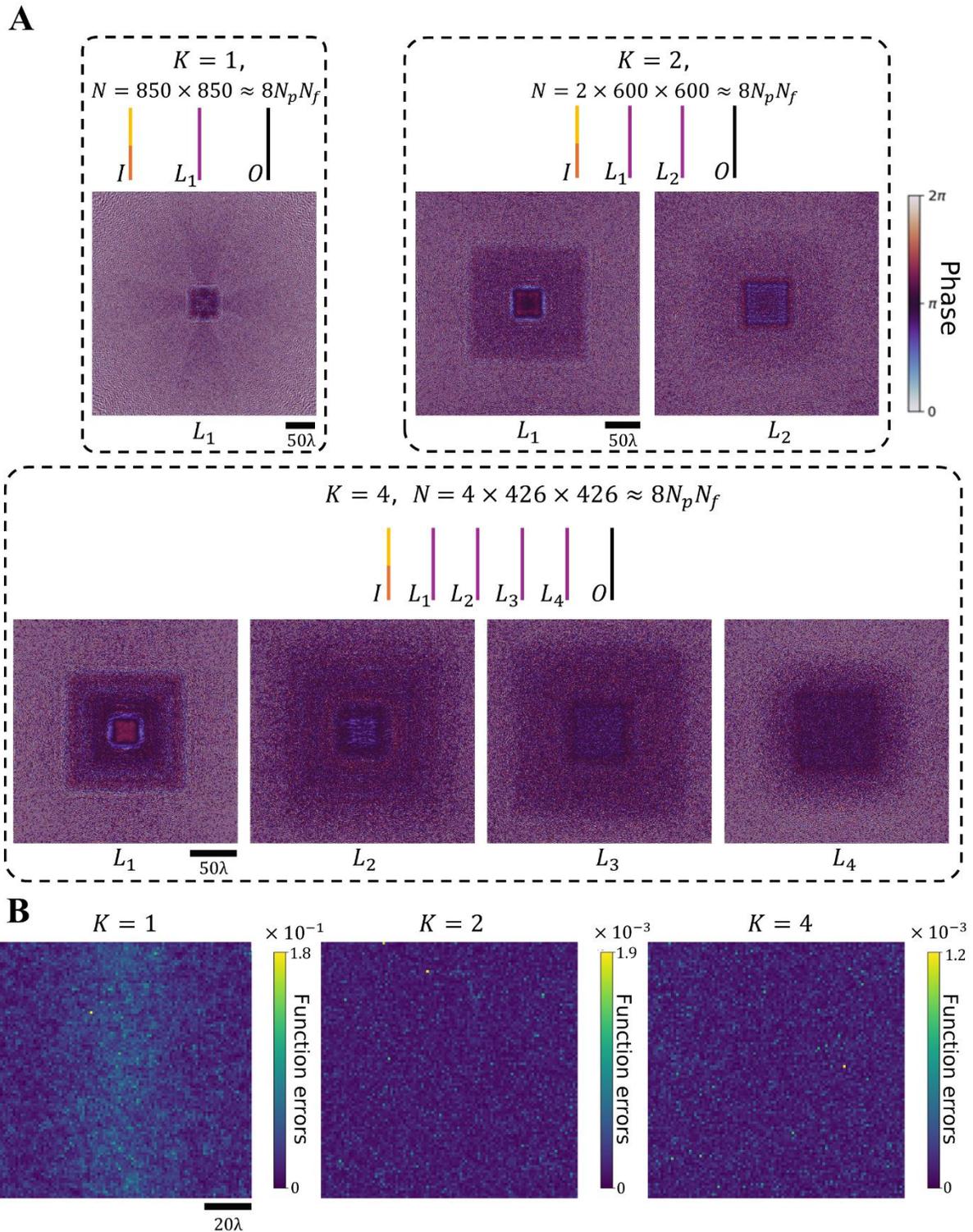

**Fig. 3 Optimized diffractive phase profiles and spatial maps of per-function approximation errors.** (**A**) Phase profiles of spatially incoherent diffractive processors with $K = 1, 2$, and $4$ diffractive layers. To ensure a fair comparison, the total number of trainable diffractive features is matched across designs ($N \approx 8 N_p N_f$), with $N$ distributed evenly over the $K$ surfaces for each design. (**B**) Spatial maps of per-function approximation error values for the corresponding designs,

showing that multi-layer diffractive processors ($K = 2$ and $4$) achieve substantially lower nonlinear function approximation errors than a single-layer design ($K = 1$) under the same number of trainable features.

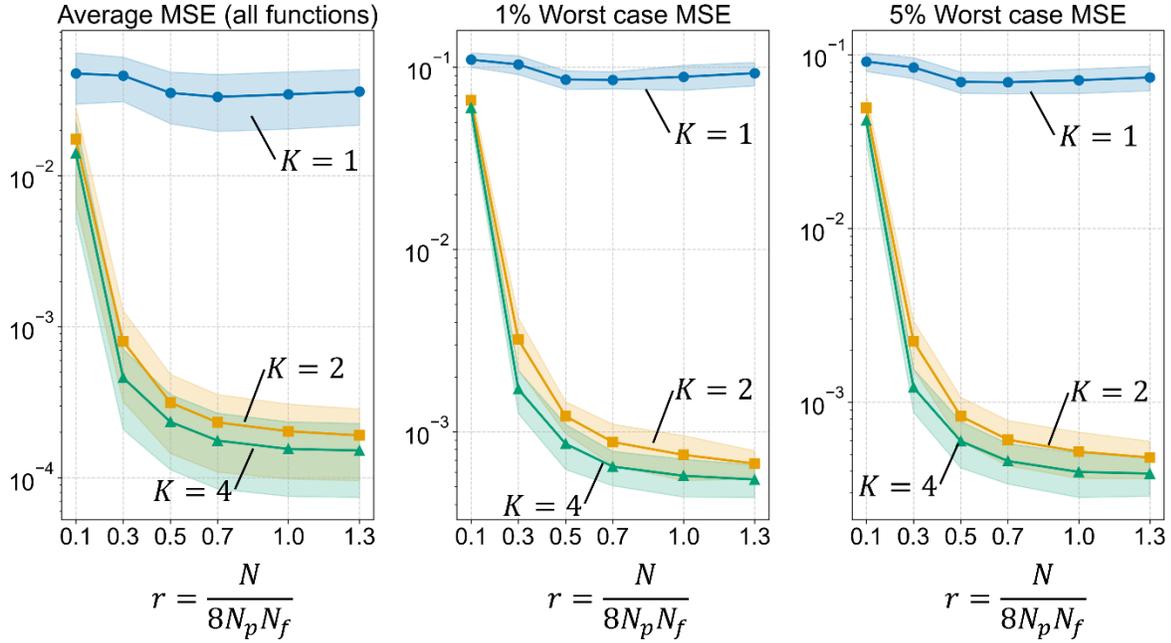

**Fig. 4 Evaluation of average per-function MSE values with respect to the number of trainable features ($N$) and diffractive depth.** Average per-function MSE (left) and tail-error metrics (middle/right; worst-case among the highest-error 1% and 5% of the functions) as a function of $r = N/(8N_p N_f)$ for $K = 1, 2,$ and $4$ diffractive layers. Increasing the number of trainable features improves both the average accuracy and tail robustness, with deeper designs ($K > 1$) consistently outperforming a single layer.

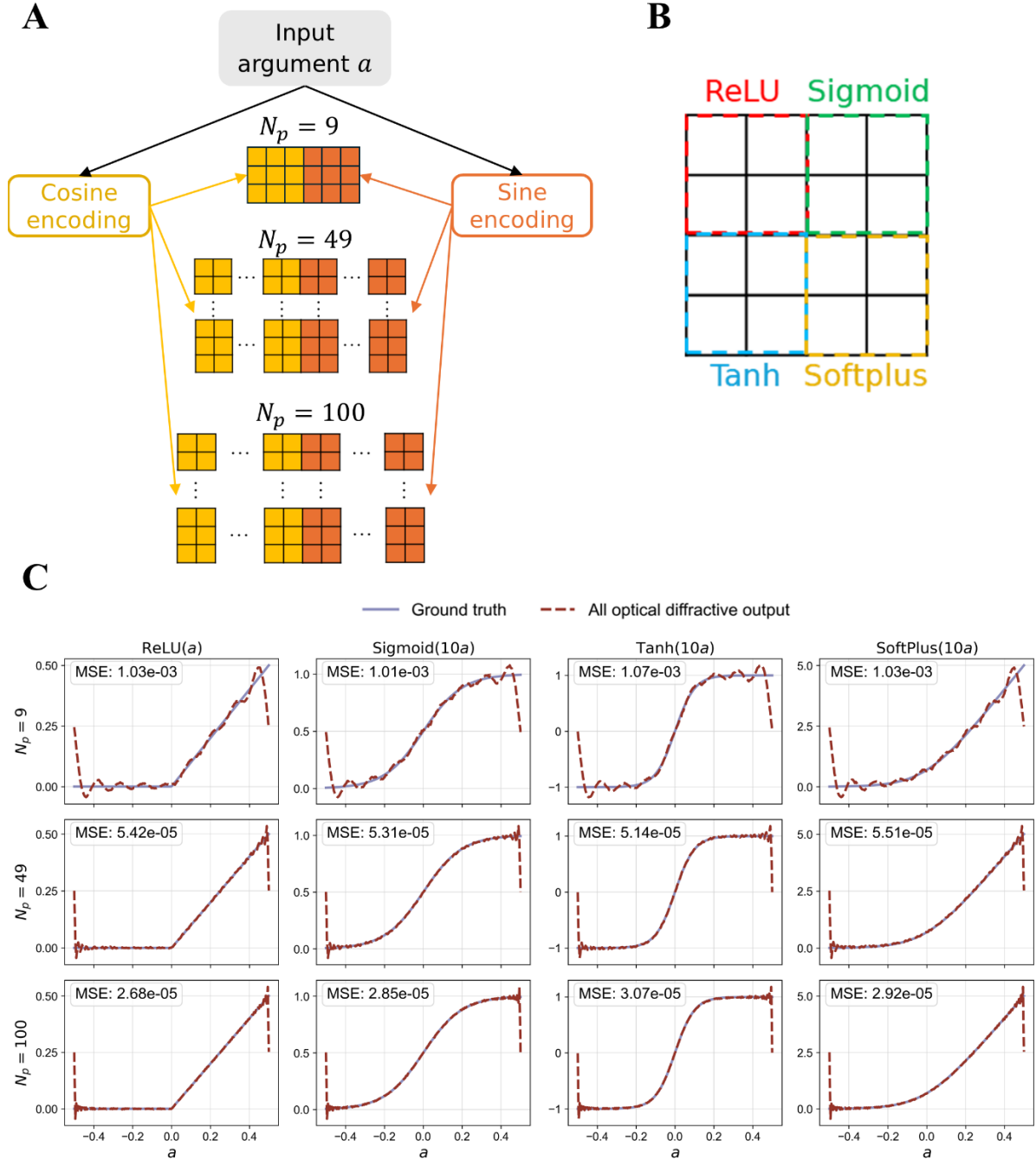

**Fig. 5 Approximation of commonly used nonlinear activation functions.** **(A)** Input-plane cosine/sine intensity encoding of the scalar argument $a$ using different encoding bandwidths $N_p$. **(B)** Output-plane multiplexing layout for the parallel evaluation of four target nonlinear functions (ReLU, Sigmoid, Tanh, and Softplus). **(C)** Approximations of the four activation functions at $N_p = 9, 49,$ and $100$. Solid curves denote the ground truth, and the dashed curves denote spatially incoherent diffractive outputs; the MSE values quantify the improved function approximation accuracy with increasing $N_p$. To accommodate different output ranges, target functions are normalized to $[0, 1]$ for training and testing as well as for the MSE computation; curves are then

rescaled to their original ranges for visualization. MSE is computed over $a \in [-0.5 + \epsilon, 0.5 - \epsilon]$ with $\epsilon = 2.5\%$ to mitigate Gibbs phenomenon-related boundary artifacts in the error estimate. Also see Supplementary Fig. 1.

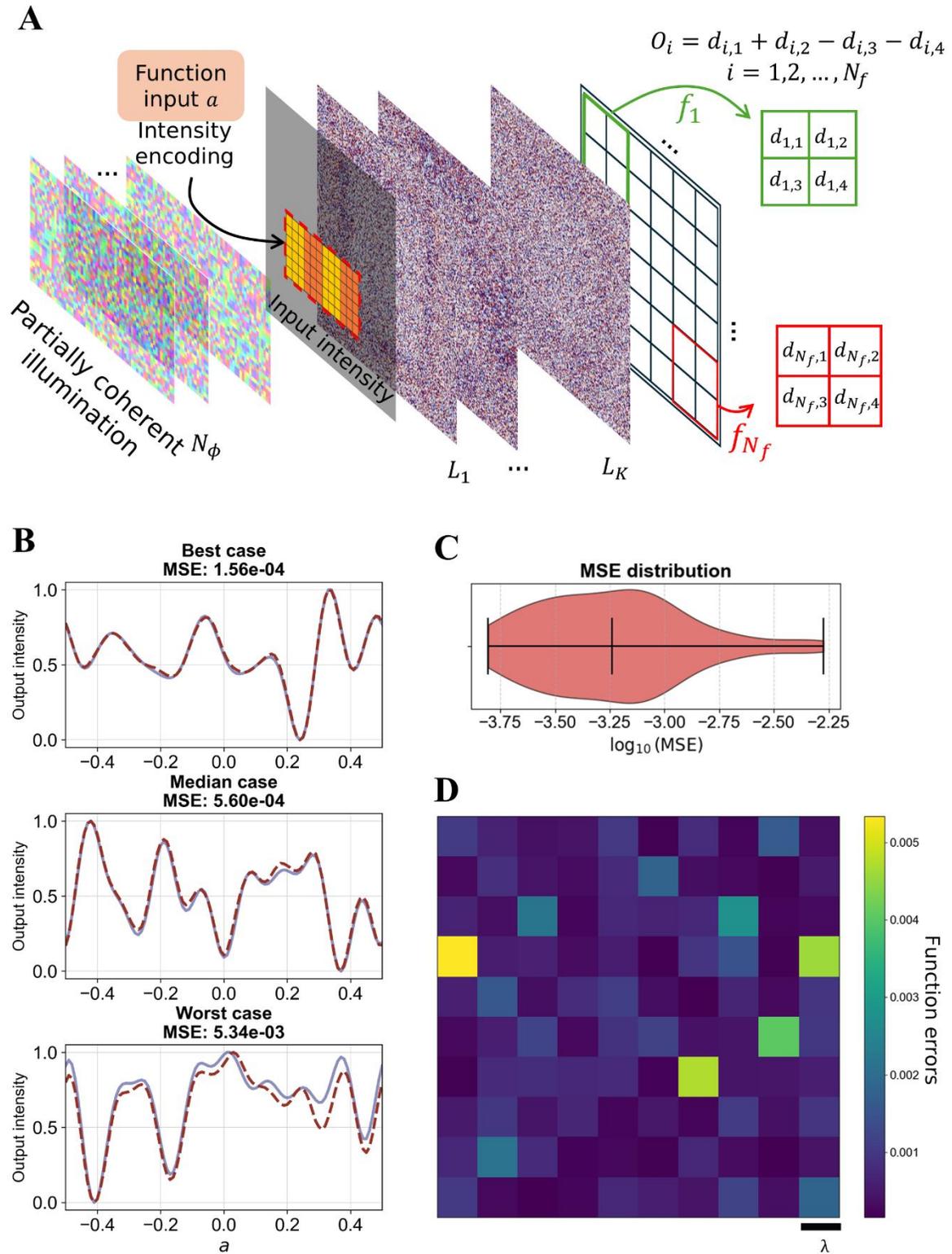

**Fig. 6 Massively parallel nonlinear function approximation with a diffractive processor under partially coherent illumination.** (A) Conceptual overview of the partially coherent diffractive processor framework. Partial coherence is modeled by averaging output intensities over

a finite number of independent random-phase realizations ($N_\phi$), after intensity-only encoding of the input argument $a$ and optical propagation through a $K$-layer diffractive processor. Each function output is decoded from a $2 \times 2$ detector tile using a differential combination of the measured intensities. **(B)** Representative approximations for the best, median, and worst cases. Solid curves denote ground truth, and the dashed curves denote the partially coherent diffractive outputs. **(C)** Distribution of per-function MSE values across all the target nonlinear functions ($N_f = 100$) under partially coherent illumination. **(D)** Spatial map of per-function approximation error values across the output plane. $K = 4$, $N_p = 9$ and $N = 4 \times 200 \times 200$ were used in this partially coherent diffractive design.

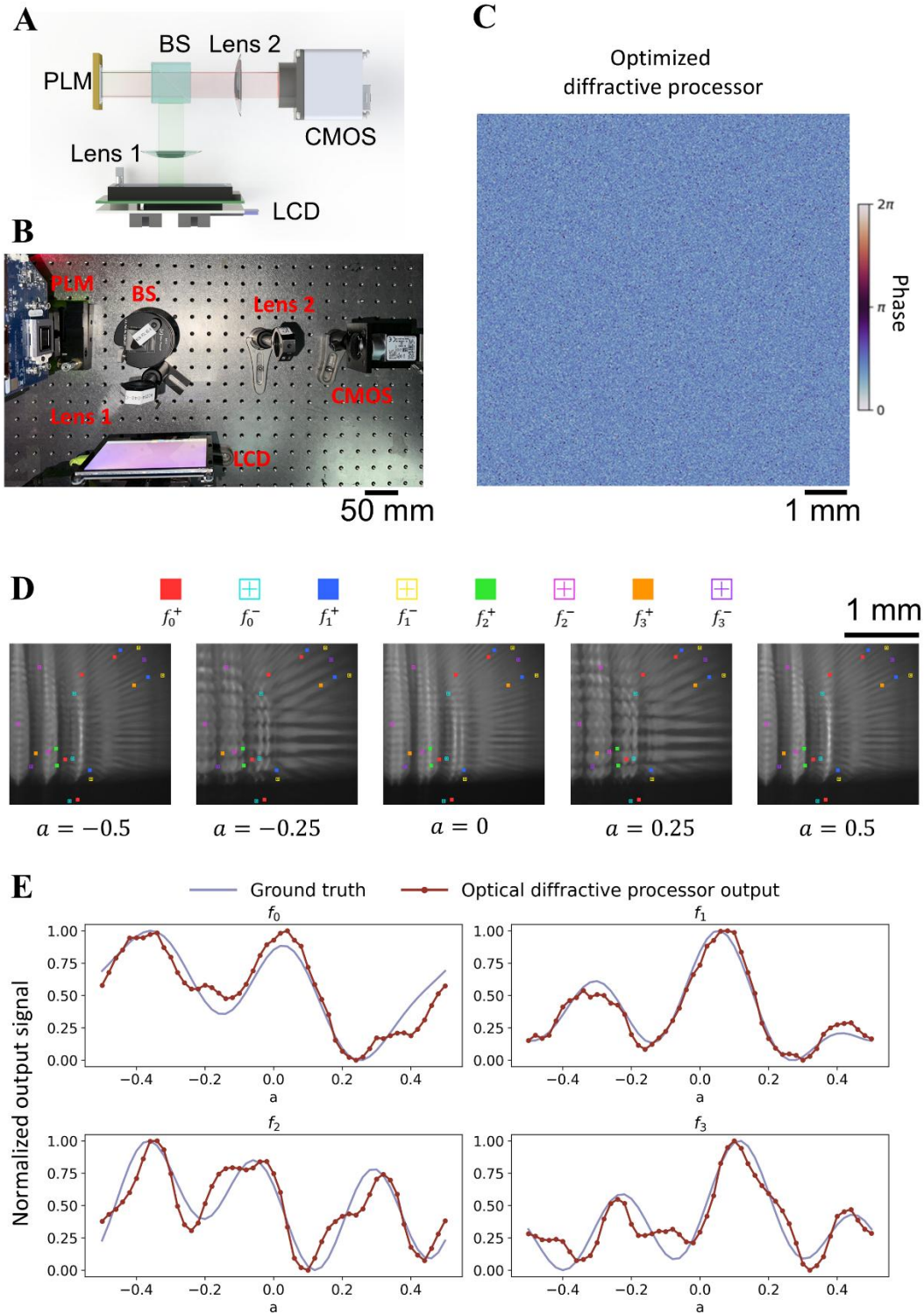

**Fig. 7. Experimental demonstration of nonlinear function approximation using incoherent LCD illumination.** **(A)** 3D schematic of the LCD-based diffractive processor for nonlinear function approximation. **(B)** Photograph of the physical experimental apparatus, comprising an LCD for input encoding, a PLM for phase modulation, a CMOS camera, and conjugate plane relay lenses. **(C)** The optimized 2D phase profile deployed on the PLM after *in situ* training. **(D)**

Representative intensity distributions captured at the CMOS image plane. Colored markers designate the final optimal spatial coordinates of the specific positive and negative detector regions that resulted from the combinatorial search for different target nonlinear functions. **(E)** Quantitative comparison between the ground truth target nonlinear functions and the experimental optical diffractive outputs ($f_0$ through $f_3$). The average per-function MSE is $1.06 \times 10^{-2}$. For visualization purposes, the raw camera images were contrast-enhanced using a linear percentile stretch (1st–99th percentile) within the original intensity range; all quantitative analyses were performed on the unprocessed raw image data.

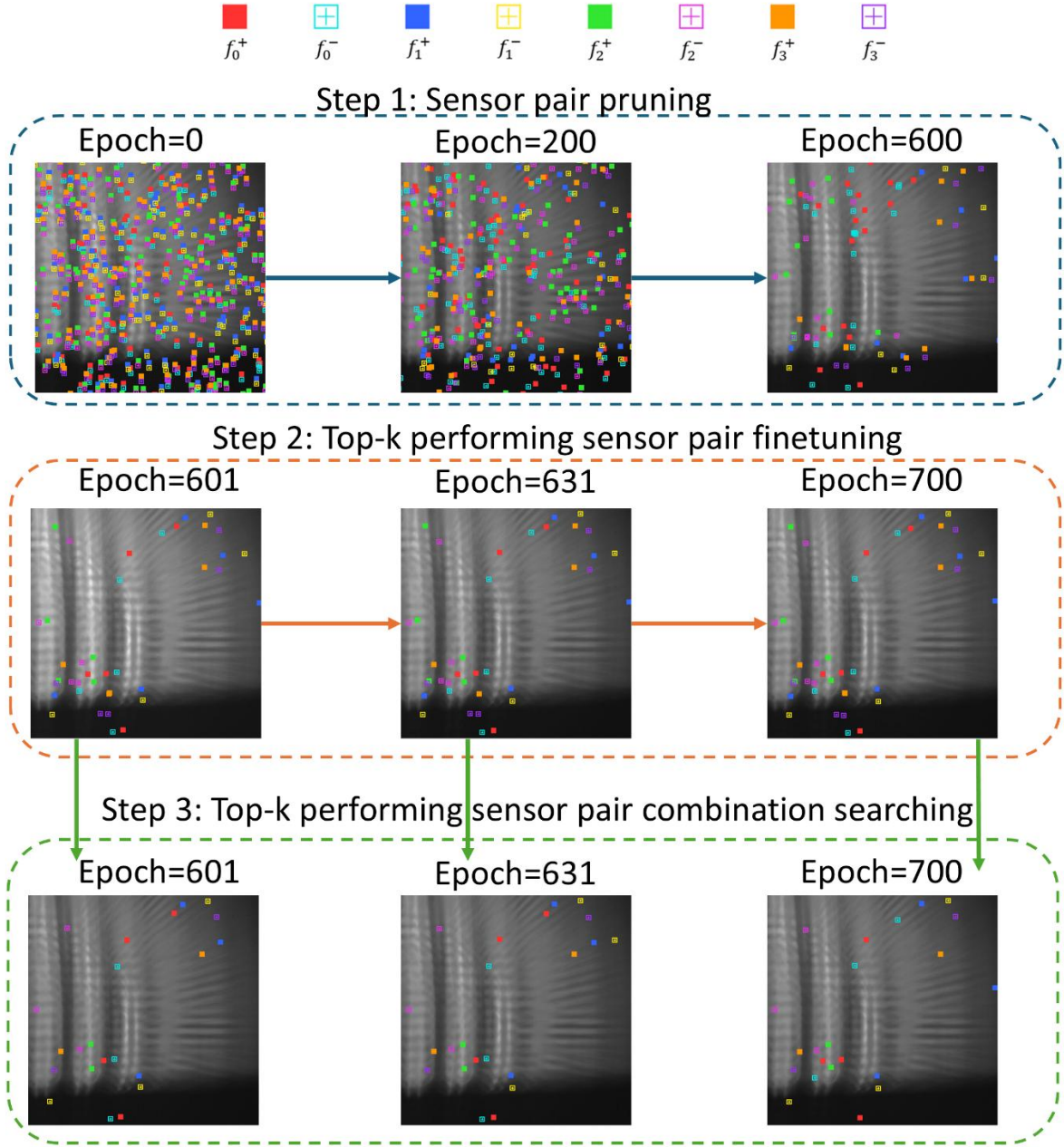

**Fig. 8. Workflow of the presented adaptive detector-pruned *in situ* learning strategy that is model-free.** We used a model-free *in situ* learning pipeline to optimize the optical diffractive processor by jointly refining the phase profile and detector coordinates directly on hardware. The top panel (blue dashed box) illustrates **Step 1: Training and pruning**, depicting the transition from an initial random distribution of 200 differential detector pairs to a sparse configuration by systematically discarding underperforming detector pairs using model-free *in situ* learning. The middle panel (orange dashed box) shows **Step 2: Fine-tuning**, where the top $k = 5$ performing sensor pairs per function are isolated for an additional 100 epochs of joint optimization using model-free *in situ* learning. The bottom panel (green dashed box) details **Step 3: Combinatorial searching**, executed over the fine-tuning trajectory. Retained detector pairs are decoupled into individual positive and negative sensor regions, and an exhaustive search assigns an optimal

inclusion weight from the set $\{-1, 0, 1\}$ to each detector region. This final optimization stage yields the highly sparse final detector configuration. For visualization purposes, the raw camera images were contrast-enhanced using a linear percentile stretch (1st–99th percentile) within the original intensity range; all quantitative analyses were performed on the unprocessed raw image data.